# Disorder induced power-law gaps in an insulator-metal Mott transition

Zhenyu Wang[1], Yoshinori Okada[2], Jared O'Neal[3], Wenwen Zhou[4], Daniel Walkup[5], Chetan Dhital[6], Tom Hogan[7], Patrick Clancy[8], Young-June Kim[8], Y. F. Hu[9], Luiz H. Santos[1,10], Stephen D. Wilson[7], Nandini Trivedi[11] and Vidya Madhavan[1, §]

[1]Department of Physics and Frederick Seitz Materials Research Laboratory, University of Illinois Urbana-Champaign, Urbana, Illinois 61801, USA

[2] Quantum Materials Science Unit, Okinawa Institute of Science and Technology Graduate University, Okinawa 904-0495, Japan

[3]Mathematics Department, The Ohio State University, Columbus, OH 43210, USA

[4]Department of Physics, Boston College, Chestnut Hill, Massachusetts 02467, USA

[5]Center for Nanoscale Science and Technology, National Institute of Standards and Technology, Gaithersburg, MD 20899

[6]Department of Physics, Kennesaw State University, Marietta, GA 30060, USA

[7]Materials Department, University of California, Santa Barbara, California 93106, USA

[8]Department of Physics, University of Toronto, Toronto, Ontario M5S 1A7, Canada

[9]Canadian Light Source, Saskatoon, Saskatchewan, Canada S7N 2V3

[10]Institute for condensed Matter Theory, University of Illinois Urbana-Champaign, Urbana, IL 61801, USA

[11]Physics Department, The Ohio State University, Columbus, OH 43210, USA

[§]Corresponding author. Email:vm1@illinois.edu

# Abstract

A correlated material in the vicinity of an insulator-metal transition (IMT) exhibits rich phenomenology and variety of interesting phases. A common avenue to induce IMTs in Mott insulators is doping, which inevitably leads to disorder. While disorder is well known to create electronic inhomogeneity, recent theoretical studies have indicated that it may play an unexpected and much more profound role in controlling the properties of Mott systems. Theory predicts that disorder might play a role in driving a Mott insulator across an IMT, with the emergent metallic state hosting a power law suppression of the density of states (with exponent close to 1; V-shaped gap) centered at the Fermi energy. Such V-shaped gaps have been observed in Mott systems but their origins are as yet unknown. To investigate this, we use scanning tunneling microscopy and spectroscopy to study isovalent Ru substitutions in $Sr_3(Ir_{1-x}Ru_x)_2O_7$ ($0 \leq x \leq 0.5$) which drives the system into an antiferromagnetic, metallic state. Our experiments reveal that many core features of the IMT such as power law density of states, pinning of the Fermi energy with increasing disorder, and persistence of antiferromagnetism can be understood as universal features of a disordered Mott system near an IMT and suggest that V-shaped gaps may be an inevitable consequence of disorder in doped Mott insulators.

Metal-insulator transitions are observed in a range of material systems from uncorrelated metals to correlated insulators[1-3]. One could start from a non-interacting `good' metal and turn it into an insulator by adding disorder as in an Anderson transition[4]. The single particle excitations in the resulting insulator remain gapless. One could also start with an interacting system such as a correlated insulator with a well-defined gap in the single particle density of states (DOS) as described by the Mott-Hubbard model[5] and turn it into a `bad' metal by doping. In the later scenario, disorder is an inevitable byproduct of doping and the interplay between correlations[2,6], doping[7] and disorder pose fundamental challenges for theory and experiment. While it is well established that disorder can lead to a soft gap in the DOS of metals[8] or insulators[9,10] with long-range coulomb interactions, it is only recently that theoretical studies have addressed the effect of disorder on the DOS of Mott systems,[11-18] with intriguing predictions of insulator-metal transitions (IMT) and a power-law (V-shaped) suppression of the density of states on the metallic side. Experimentally, disorder is well known to create an inhomogeneous potential landscape. However, whether disorder might lead to power law gaps in doped Mott systems, remains a key outstanding question that has not yet been experimentally addressed. This is a particularly important question since such power law suppressions of density of states are ubiquitously observed in doped Mott systems (supplementary section (SI) I), but their origins remain a mystery.

The iridium oxide compounds (iridates) provide an ideal platform to address the effect of disorder in doped Mott systems. The parent compounds of the layered iridate $Sr_{n+1}Ir_nO_{3n+1}$ (n=1, 2) host novel $J_{eff}$=1/2 Mott ground states, where strong spin-orbit coupling and crystal field effects split the $5d^5$ $t_{2g}$ manifold into an occupied $J_{eff}$ =3/2 and a narrow half-filled $J_{eff}$ =1/2 band, which in turn amplifies correlation effects and results in a Mott-antiferromagnetic (AFM) insulator[19,20]. Electron doping these materials has been shown to result in intriguing phenomena which parallel the hole-doped cuprates[21,22] Among the iridates, pristine $Sr_3Ir_2O_7$ (Ir327) exhibits a smaller charge gap (ΔE~130meV) and is ideally suited to scanning tunneling microscopy (STM) studies[23].

In this work we focus on Ru-doped Ir327. One of the biggest problems in studying disorder effects experimentally is that doped Mott systems often host a variety of competing phases, which makes it extremely difficult to isolate the effects of disorder. Figure 1A depicts the electronic phase diagram of $Sr_3(Ir_{1-x}Ru_x)_2O_7$ which has been well established in our previous study[24]. As seen in the phase diagram, other than AFM, there is a distinct absence of additional order parameters. In addition, in current study, we constrain ourselves to Ru concentrations far from the paramagnetic phase boundary in the region where static AFM is robust and AFM fluctuations play a minimal role. Furthermore, STM data show that Ru acts as a weak scatterer and unlike most other impurities does not induce in-gap impurity states. The lack of competing phases, and the role of Ru as a weak perturbant make $Sr_3(Ir_{1-x}Ru_x)_2O_7$ an ideal system to study the effects of disorder on the DOS in Mott insulators.

While 5% of electron doping[25] or 15% of hole doping[26] drives the Mott transition, up to 37% in-plane Ru substitution is required to collapse the parent Mott state (Fig. 1B). These resistivity data are consistent with recent x-ray absorption spectroscopy[27] (XAS) and resonant inelastic x-ray scattering data[28] which show that Ir retains a formal valence of 4+ even with high Ru substitution and implies that the electronic configuration of Ir remains unchanged with Ru substitution. This can be contrasted with the case of Rh doping[26], where Rh clearly introduces holes into the system and correspondingly changes the valence state of Ir from 4+ to a mixture between 4+ and 5+. The difference between Ru and Rh can be explained by our data which show that unlike Rh which has a valance of 3+ in Ir327[29], Ru goes into the lattice with a 4+ valence state identical to Ir (Fig. 1C). Taken together with our STM measurements where the Fermi energy does not move into the valence band with increasing Ru doping, we conclude that the Ru does not significantly change the itinerant carrier concentrations at these dopings, which explains the large amount of Ru needed to make the system metallic.

Ir327 is composed of alternating $IrO_6$ bilayers separated by rock-salt SrO spacers. Cleavage occurs easily between two adjacent SrO layers, resulting in a charge-balanced surface. Figure 1D depicts a typical STM topograph obtained on a x=0.5 sample. The Sr atoms form a square lattice with lattice constant a=3.9Å. It is difficult to see the individual Ru atoms in this image because Ru substitutes on the Ir sites, which lie 2.0 Å below the SrO plane. However, at higher Ru concentrations we observe patches of bright regions in the topography, which as we will see later represent areas with a smaller gap magnitude caused by the Ru substitution. A comparison of the topographies for samples with different Ru doping levels can be found in SI II.

Local electronic structure can be probed by a differential conductance (dI/dV) measurement which is proportional to the local DOS. We start with the parent compound which has been discussed in detail in our earlier paper[23]. The spatially averaged *dI/dV* spectrum obtained on defect-free regions of the undoped sample is shown in SI III. The DOS in an energy range from -10meV to +120meV is suppressed to zero, resulting in a hard Mott gap ~130meV with Fermi level positioned close to the top of the lower Hubbard band. To study the effect of Ru substitution, we first examine spatially resolved *dI/dV* spectra collected along a line (linecut) in samples with varying Ru concentration across the IMT. At x~27% Ru doping, transport properties indicate an insulating behavior, with neutron scattering showing robust antiferromagnetism, confirming that the compound is still in the Mott phase. Figure 2A displays a linecut across dark and bright regions of the topography at this doping. The tunneling spectra show pronounced spatial variations: in the dark region, the DOS is zero for the energy range from -10 to +100 meV, yielding a gap size of ~110 meV; This gap becomes much narrower when approaching a bright region. Although the connection between Ru substitutions and bright areas is not at first obvious, the local electronic structure shows a clear trend with increasing Ru doping. In the x~35% sample which lies near the IMT (Fig. 2B), the insulating gap obtained in the dark areas diminishes (~80meV), and even vanishes in the bright area, changing into a V-shaped gap. Note that the term `V-shape' is used in this paper to describe the power law (almost linear) dependence of the density of states on energy. This particular spectral shape will be discussed in further detail later. The sample with x~50% is on the metallic side of the IMT and the linecut shown in Fig. 2C reveals that spatial inhomogeneity persists into this nominal metal. However, other than some extremely Ru-rich regions where the spectra are similar to that of $Sr_3Ru_2O_7$ (ref.30), the data now predominantly show a V-shaped gap.

While spectra in line cuts provide a glimpse into the evolving spectral features across the IMT, to obtain a more complete picture, we acquire dI/dV (r, V) spectra on a densely-spaced 2D grid (dI/dV map) for each doping (SI IV). The magnitude of the charge gap is then extracted at each location in the grid (SI V) and we plot the 2D gapmaps in Fig. 3A-C. To understand the range of spectra within each sample and to see how the spectral shapes evolve with doping, we sort the spectra for each map into bins defined by the gap magnitude, and display the averaged spectrum for each bin (Fig. 3E-H). Further details of the sorting procedure are in SI VI. The gap maps are color coded to match the spectra; i.e., red/blue/green areas of the map for a particular doping, correspond to the red/blue/green *dI/dV* spectral shapes shown for that doping.

At x~27%, consistent with the insulating behavior of the resistivity, the sample shows gapped spectra in most areas (Fig. 3A and E) with an average gap of 72meV (as shown by the histogram in Fig. 3I). Very close to the IMT, at x~35%, the gap histogram shifts to lower energies and the average gap is ~40 meV. Clear V-shaped gap features can now be seen that cover about 20% of the sample (red areas). We associate these V-shaped areas with metallic regions. Eventually at the IMT at x~37.5% (Fig. 3C and G), the regions showing V-shaped spectra become predominant. The metallic regions have grown in spatial extent and now mutually connect (red and yellow in the gapmap), in agreement with a quantum percolation driven IMT transition proposed in transport studies[31]. Finally, we turn to the metallic x~50% compound. To obtain a visual sense of the variation of the spectra in this gapless 50% sample, a *dI/dV* conductance map at 30meV is plotted in Fig. 3D and the associated spectra are shown in Fig. 3H. We see that the sample on the metallic side continues to remain inhomogeneous. The metallicity is represented by the pervasive V-shape of the spectra, and the inhomogeneity is represented by the variation in spectral lineshapes. We note here that while the spectral shapes differ with location and doping, the Fermi level is always pinned close to the top of lower Hubbard band.

**Discussion and Summary**

The question we seek to address is: how do we understand striking V-shaped gap observed in $Sr_3(Ir_{1-x}Ru_x)_2O_7$? Similar gaps have been seen in other Mott systems and are often attributed to the emergence or fluctuations of an order parameter. This is however an unlikely explanation for the V-shaped gap seen in our data. Unlike hole doped cuprates[32, 33] and the electron doped single layer iridates[34], and consistent with the phase diagram established in earlier studies, no sign of additional charge order is observed either in real or momentum space in this system (SI VIII). Combined with the absence of superconductivity in this phase diagram, we rule out new order parameters or preformed pairs as the cause of this gap. Another possibility suggested by calculations is that the soft gap arises from AFM fluctuations. However, in our case, it is clear from neutron scattering data[24] that static antiferromagnetism persists deep into the metallic phase where the V-shape gap is pervasive. In fact, the phase transition into the paramagnetic state occurs at much higher percentages of Ru ~65%. Since we observe a pervasive V-shaped gap in the metallic sample with robust long-range AFM, it is unlikely that AFM fluctuations are the cause of this gap.

To explain this phenomenology, we turn to the disordered Hubbard model, which incorporates disorder as a random site dependent potential variation. While this may be a simplistic model for our system, it is fruitful to compare the theoretical results with our experimental data to look for universal signatures of a disordered Mott system. The disordered Hubbard model has been numerically studied in the last decade both by Hartree-Fock[11,14] as well as quantum Monte Carlo[13] techniques (SI IX). We reproduced the Hartree-Fock calculations with a more realistic disorder potential and the results show remarkable similarities with our data (SI X). First, with increasing disorder fraction, the Mott gap gradually closes; second, the Fermi level is pinned at the top of lower Hubbard band and the Mott gap fills up by states being pulled in from the upper Hubbard band; and third, after the IMT, the calculated spectra show a characteristic V-shape similar to our data. Interestingly, the simulations also show a pronounced spatial inhomogeneity of DOS, where Mott-gapped insulating regions coexist with -V-shape gapped metallic regions (SI X). These multiple observations taken together suggest that the V-shaped gap may be attributed to the effects of disorder on the electronic structure of a Mott insulator.

To further establish the relationship between the observed gap and a disordered Mott system, we look for a universal scaling relationship of the density of states with energy ($D(\epsilon)$). We find that most of the numerical calculations using the Hubbard model predict a power-law dependence[12-14,35,36] for $D(\epsilon)$. We track the power-law exponent in the experimental data across the IMT by fitting the tunneling spectra shown in Fig. 3E-H with a power-law function $D(\epsilon) \sim |\epsilon - \epsilon_F|^p$ in energy range about [-50meV, 50meV]. The fitting parameter $p$ for each indexed spectrum is plotted in Fig. 4B. Since each sample shows a range of spectral shapes, the value of the exponent correspondingly varies. However, we see a clear trend across the IMT. For illustration, if we consider the most metallic spectra for each doping, we find that the exponent $p$ changes from a value around 2 (U-shaped) in the 27% doped samples to about 1 in the inhomogeneous metallic samples, for both the occupied and empty states, confirming the that the suppression in the density of states is quite linear at low energies (Fig. 4A and B). The change in exponent from 2 to 1 is a non-trivial occurrence in both theory and experiment. A simple band closing picture, due to either carrier doping or reduced spin-orbit coupling, would not be able to capture the linear behavior of the density of states and a deeper understanding of the origin of this power law depletion in the density of states would be useful. Interestingly, a heuristic, albeit very general, statistical model can be constructed to explain how a linear DOS might arise in a standard Anderson-Hubbard model in the highly disordered regime, which is described in the SI XI.

On comparing our findings with other Mott systems (SI I), we observe many common features including IMTs which proceed through spatial inhomogeneity and V-shaped gap formation. In SrRu$_{1-x}$Ti$_x$O$_3$, for example, which is another candidate to reveal the nature of disorder in strong correlated systems, a nearly linear energy dependence of DOS has been found[37]. Gaps with similar lineshapes have also been reported in many doped oxides including other iridates[26, 34, 38], underdoped cuprates[33, 39, 40] and manganites[41] (SI I), making it is essential to consider disorder effects in these systems. While the complex phenomenology of the cuprates and iridates cannot all be captured by disorder effects due to the existence of additional electronic orders in the phase diagram, disorder may be implicated in the linear suppression of density of states near the Fermi energy in these systems. Though how the disorder effect contributes to the formation of the pseudogap remains an interesting topic for future study, we conclude here that V-shaped gaps centered at E$_F$ in Mott insulators do not necessarily require superconductivity or additional order parameters like charge density waves.

In correlated metals, it has been shown both theoretically and experimentally that disorder can cause localization and decrease conductivity by generating a soft gap. Here we approach the problem from the opposite side of the phase diagram. Starting with a correlated insulator we demonstrate experimentally that a chemical disorder can also induce a power-law gap which does not require any additional order parameter for its existence.

## Materials and Methods

Sr327 single crystals used in our experiments were grown by the conventional flux methods[24], cleaved at ~77 K in ultra-high vacuum and immediately inserted into the STM head where they are held at ~5 K during the process of data acquisition. All dI/dV measurements were taken using a standard lock-in technique with ~5-10mV peak to peak modulation. Tungsten tips were annealed and then prepared on Cu single crystal surface before using on iridate samples.

X-ray absorption spectroscopy measurements were performed using the Soft X-ray Microcharacterization Beamline (SXRMB) at the Canadian Light Source. Measurements were carried out at the Ru L3 ($2p_{3/2}$ to 4d) and L2 ($2p_{1/2}$ to 4d) absorption edges, which occur at energies of 2838 eV and 2967 eV respectively. Data were collected using both Total Electron Yield (TEY) and Fluorescence Yield (FY) detection modes. Energy calibration was verified by a comparison of Ar K edge features observed at E = 3206 eV. In order to improve counting statistics and verify the reproducibility of the data, multiple scans were performed on each sample. Each dataset shown in Fig. 1C or Fig. S2 consists of three or more individual scans which have been binned together. Each scan has been normalized such that the edge jump associated with the Ru absorption edge has been set to unity.

## Acknowledgements

We sincerely thank Steve Kivelson, Ziqiang Wang, Peter Abbamonte, Nadya Mason, and Eduardo Fradkin for useful conversations. V.M. gratefully acknowledges funding from NSF Award No. DMR-1610143 for the STM studies. The theoretical collaboration was supported by DMREF 1629382 and NSF DMREF 1629068 and in part by the Gordon and Betty Moore Foundation's EPiQS Initiative through the Grant No. GBMF4305 at the Institute for Condensed Matter Theory of the University of Illinois (L. S). S.D.W. acknowledges funding support from NSF Award No. DMR-1505549. Research at the University of Toronto was supported by the NSERC, CFI, OMRI, and Canada Research Chair program. We used the Canadian Light Source, which is funded by the Canada Foundation for Innovation, the Natural Sciences and Engineering Research Council of Canada, the National Research Council Canada, the Canadian Institutes of Health Research, the Government of Saskatchewan, Western Economic Diversification Canada, and the University of Saskatchewan.

# Figure 1

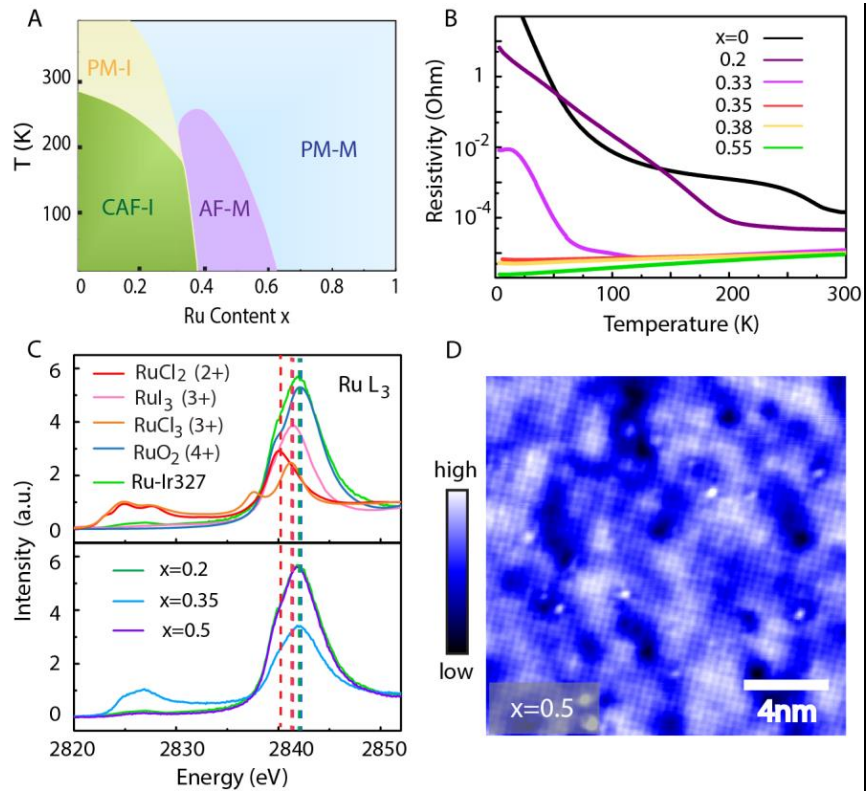

**Figure 1.** Insulator-metal transition in $Sr_3(Ir_{1-x}Ru_x)_2O_7$. (A) Phase diagram of $Sr_3(Ir_{1-x}Ru_x)_2O_7$ established by bulk susceptibility, neutron scattering and transport measurements in ref. 33. CAF-I: insulating canted AF phase; PM-I: paramagnetic insulating phase; AF-M: AF ordered metallic phase; PM-M: paramagnetic metal. (B) Temperature-dependent resistivity for different Ru concentrations. IMT takes place at the critical concentration of x~0.37. (C) Comparison of the Ru oxidation state in $Sr_3(Ir_{1-x}Ru_x)_2O_7$ with other reference samples using X-ray absorption spectroscopy. The oxidation states of Ru in $RuCl_2$, $RuI_3$, $RuCl_3$ and $RuO_4$ are 2+, 3+, 3+ and 4+ respectively. (D) Topography taken on 50% Ru-doped samples. Scanning conditions: $V_b$=200mV, I=50pA. $V_b$: sample bias with respect to the tip; I: set point current.

**Figure 2**

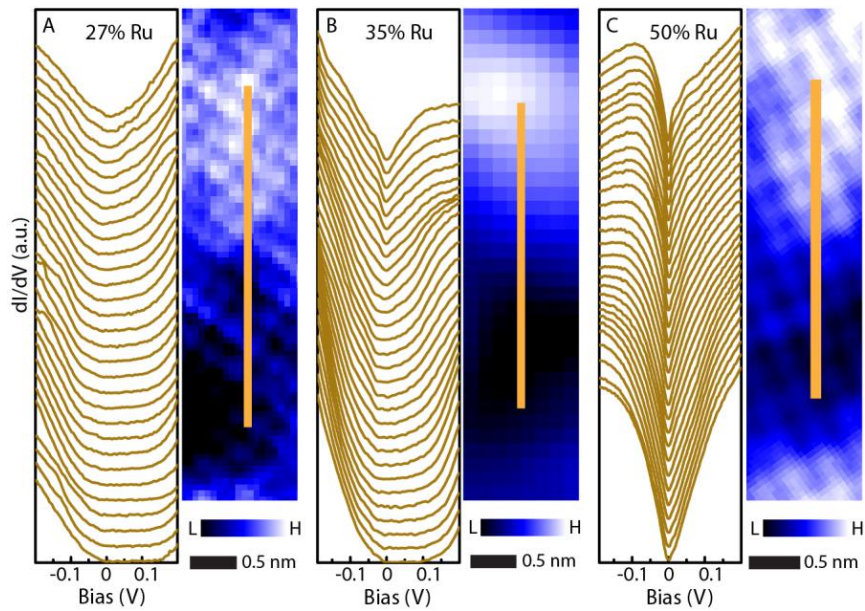

**Figure 2.** *dI/dV* line cuts across IMT in $Sr_3(Ir_{1-x}Ru_x)_2O_7$. Line cuts and the associated topographic images indicating where the line cuts were obtained in: (A) 27%, (B) 35%, and (C) 50% Ru-substituted samples. The spectra in A and B show the evolution from gapped and insulating line-shape (27%), to relatively smaller gap and more metallic behavior (35%). The spectra of 50% Ru-doped sample shown in C, however, exhibit V-shaped DOS almost along the whole line. STM setup condition in all panels: $V_b$=-200mV, I=100pA. The spectra have been shifted vertically for clarity.

.

# Figure 3

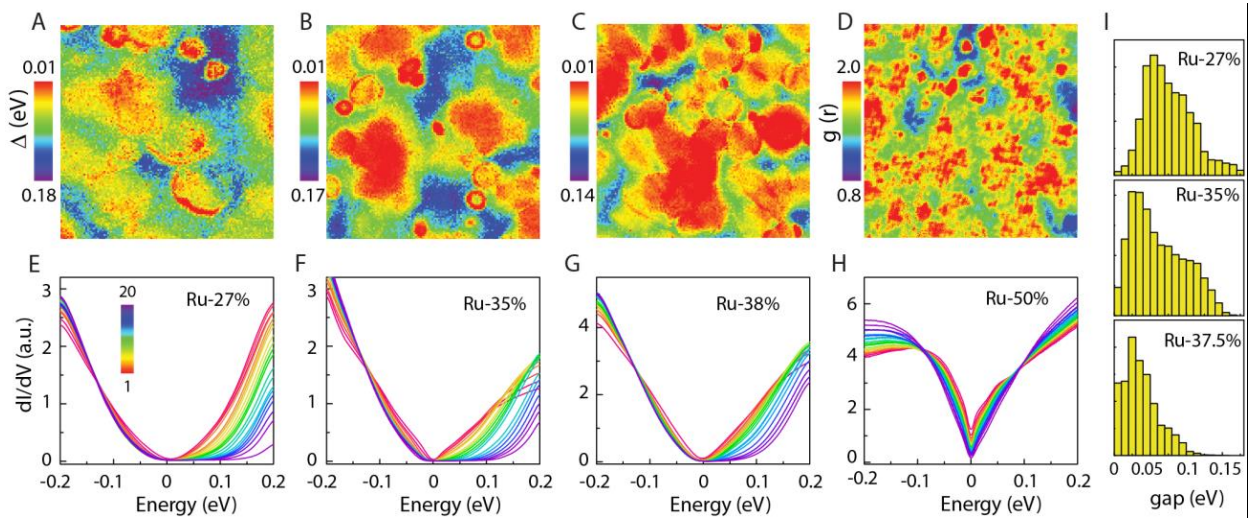

**Figure 3.** Spatial evolution of *dI/dV* spectra for a wide range of Ru substitution. (A-C) Gap maps showing the spatial inhomogeneity. The metallic regions expand (red) as the Ru concentration increases, indicating the trend towards an IMT. (D) *dI/dV* conductance map at +30mV for x=0.5 Ru- substituted sample. (E-G) Averaged *dI/dV* spectra for x~27%, x~35% and x~38%, respectively. To create these averages, spectra in the dI/dV maps were sorted by the gap magnitude into bins and then each bin was averaged. The spectra were split into ~20 bins with equal population, ranging from the ones with the largest gap (blue/purple) to the ones with zero gap (red). The colors for the spectra at each doping depict the spectra in the same color regions of the gap maps at that doping. (H) Averaged spectra which are classified by the conductance value at +30mV. At this Ru concentration, all spectra are gapless. The color scale indicates the conductance value (see section VIII). (I) Histograms of the spectral gap magnitude. The average gap size shrinks as the Ru substitution increases.

# Figure 4

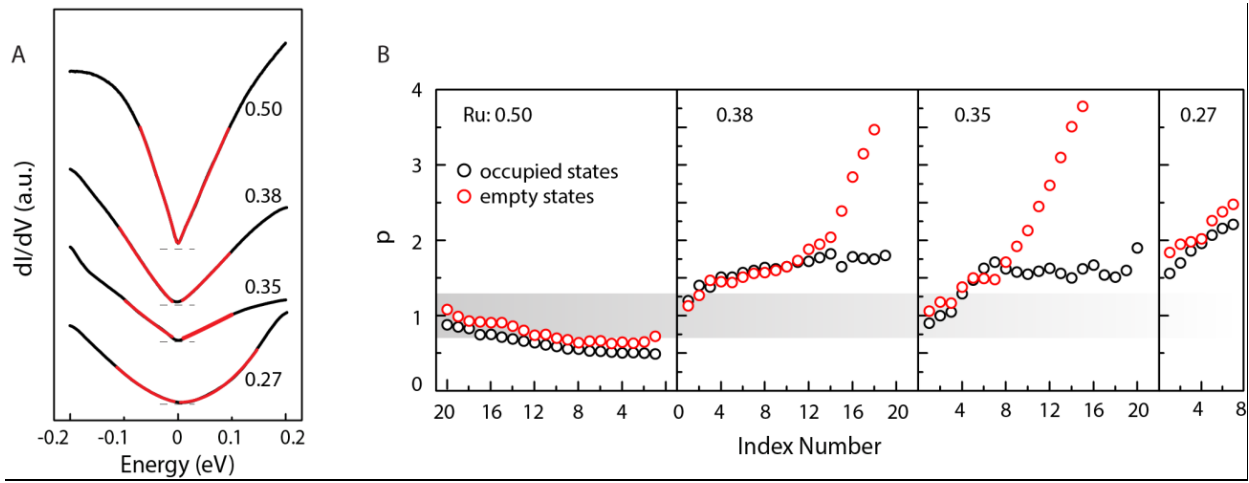

**Figure 4.** Characterizing the V-shape gap. (A) Fit to the most metallic dI/dV spectra shown in Fig. 3 for x=0.27, 0.35 and 0.38 , and the most V- shaped one in Ru=0.5, with $a|(E-E_F)|^P$. The spectra have been shifted vertically for clarity, and the dashed lines denote the position of zero conductance for each curves. (B) Range of fitting parameter *p* for the data shown in Fig. 3E-H. The index numbers label the bins described in Fig. 3. A smaller power corresponds to a more metallic spectrum.